\documentclass{ifacconf}

\usepackage{graphicx}
\usepackage[caption=false]{subfig}
\usepackage{natbib}

\usepackage{amsmath}
\usepackage{amssymb}
\usepackage{mathtools}
\usepackage{bm}
\usepackage{mathrsfs}
\usepackage{accents}

\usepackage{standalone}
\usepackage{tikz}
\usetikzlibrary{arrows.meta, positioning}

\newcommand{\matr}[1]{\mathbf{\MakeUppercase{#1}}}              
\newcommand{\vect}[1]{\mathbf{\MakeLowercase{#1}}}              
\newcommand{\est}[1]{\widehat{\mathbf{\MakeLowercase{#1}}}}     
\newcommand{\Ehat}{\widehat{\mathbb{E}}}                        
\newcommand{\cond}{\,|\,}                                       

\usepackage{comment}

\makeatletter
\let\ifac@ssect\@ssect
\makeatother
\usepackage[hidelinks]{hyperref}
\makeatletter
\def\@ssect#1#2#3#4#5#6{%
  \NR@gettitle{#6}%
  \ifac@ssect{#1}{#2}{#3}{#4}{#5}{#6}}
\makeatother

\begin{document}
\begin{frontmatter}

\title{Do Hierarchical Networks Actually Learn?\\
Insights from Word-of-Mouth Estimation}

\thanks[footnoteinfo]{This work was supported by the Swedish Research Council under contract number 2023-05170, by the Wallenberg AI, Autonomous Systems and Software Program (WASP), and by the Swedish Civil Defence and Resilience Agency (Project MAD-VAMCHS).}

\author[First]{Andrea Da Col}
\author[First]{Cristian R. Rojas}

\address[First]{%
Department of Decision and Control Systems, \\
KTH Royal Institute of Technology,
100 44 Stockholm, Sweden,
(e-mails: andreadc@kth.se, crro@kth.se).
}

\begin{abstract}
We study recursive estimation of a constant state vector by a hierarchical network of asymmetrically informed linear estimators, modeled as a directed tree. 
Only the leaves observe the regression signal, whereas every other estimator learns by \emph{word of mouth} from noisy signal estimates relayed by its children. 
Authority within the hierarchy is expressed through common prior information: after each round, every estimator adopts the estimate and error covariance broadcast by the root. 
The more accurate this common information, the larger the effective noise variance on local observations, which makes estimation intrinsically self-defeating. 
The resulting performance is the subject of this paper. 
We show that recursive least-squares (RLS) estimators do learn the state in mean square, albeit slowly. 
Under strong persistency of excitation, the broadcast error covariance decays polynomially in time, at a rate that we characterize by upper and lower bounds of matching order. 
The exponent is determined by the smallest number of relays between a leaf and the root, and it decreases exponentially with this number.
\end{abstract}


\end{frontmatter}

\section{Introduction}
In many estimation problems, those who measure are not those who decide.
In organizations, multi-level sensor fusion architectures, and chains of command, data are collected at the periphery, condensed into estimates, and passed upward, whereas the assessment formed at the top is sent back down and adopted by everyone.
Such hierarchies are economical in communication and simple to operate.
Yet every layer above the first learns from the estimates of the layer below, and no longer from data.
This paper asks whether, and how fast, a hierarchy of this kind can identify an unknown state vector.

From the viewpoint of system identification, the answer is not obvious.
For a single agent, RLS under persistent excitation is consistent, with an error covariance of order $t^{-1}$ \citep{lai1982least,ljung1983theory,goodwin2014adaptive}.
Networked extensions typically preserve this picture: in diffusion and consensus schemes, agents exchange information with their neighbors and attain near-centralized performance, and cooperation may even relax the excitation required of each agent \citep{olfati2007distributed,xie2018analysis}.
In the identification of dynamic networks, the interconnection is instead the object to be identified, and path-based graph conditions determine what can be identified \citep{van2013identification, hendrickx2018identifiability}.
Our setting is complementary to both these lines of work: the network is the estimator, information flows in one direction, and the estimate formed at the top overrides every local one.

The closest precedents come from social learning, where agents learn from the actions of others and may herd on them \citep{bikhchandani1992theory,chamley2004rational}.
\citet{vives1993fast} showed that rational agents who learn from a noisy public statistic of each other's actions accumulate precision at the rate $t^{1/3}$, in place of $t$: the more informative the public statistic, the less agents respond to their private signals, and the less new information it incorporates.
This slowdown is not cured by numbers, since the speed of learning from actions remains bounded irrespective of the group size \citep{harel2021rational}.
Closest to our work, \citet{krishnamurthy2024slow} studied a chain of interacting Kalman filters and proved that its learning rate degrades exponentially with the length of the chain.
Their analysis is restricted to a scalar Gaussian state, a constant regressor, and a single path.

We extend this line of work to directed trees, vector states, and time-varying regressors, in a second-order setting that requires no Gaussianity.
Our contributions are threefold.
First, we show that every local RLS update is the linear minimum mean-square error (LMMSE) estimate given the information that the protocol retains, and that the entire network is equivalent to a single RLS estimator.
Its noise variance is an effective resistance of the tree, viewed as an electrical network, and it diverges as the common estimate improves.
Second, under strong persistency of excitation, we bound the error covariance from above and below by multiples of $t^{-\beta}$, in every direction of the state space.
Since the equivalent noise variance is unbounded and coupled to the covariance itself, standard RLS arguments do not apply.
Third, we characterize $\beta$ explicitly: it equals $1/3$ with a single relay, in agreement with \citet{vives1993fast}, and is at least halved by every additional relay on a shortest directed path from a leaf to the root.
No other feature of the tree enters, neither the number of agents nor the number of branches, and for a directed path with a scalar state we recover the rate of \citet{krishnamurthy2024slow}.
In a nutshell, hierarchical networks do learn, but slowly, and it is their height, not their size, that matters.
We illustrate these findings numerically and discuss remedies based on the topology, the protocol, and forgetting.

The paper is organized as follows. 
Section~\ref{sec:preliminaries} collects preliminaries, and Section~\ref{sec:problem formulation} formulates the problem.
Section~\ref{sec:structural results} presents the structural results, and Section~\ref{sec:convergence analysis} the convergence analysis. 
Section~\ref{sec:numerical examples} provides numerical examples and a discussion, and Section~\ref{sec:conclusions} concludes the paper.
\section{Preliminaries}\label{sec:preliminaries}

\emph{Basic notation:}
We denote by $\mathbb{N}$ the set of natural numbers, not including zero, and by $\mathbb{R}$ the set of real numbers.
Scalars are represented by ordinary lowercase letters, column vectors by bold lowercase letters, and matrices by bold uppercase letters. 
We write $\mathbf{1}$ and $\matr{I}$ for the column vector of all ones and for the identity matrix, respectively, with dimensions determined by the context. 
The trace of a square matrix $\matr{A}$ is denoted by $\operatorname{tr}(\matr{A})$. 
For $\matr{A} = \matr{A}^\top$, we write $\matr{A}\succ \matr{0}$ if $\matr{A}$ is positive definite and $\matr{A}\succeq \matr{0}$ if it is positive semidefinite.

\emph{Random variables and orthogonal projections:}
Let $(\Omega, \mathcal{F}, \mathbb{P})$ be a probability space, and let $\mathbb{E}$ denote expectation with respect to $\mathbb{P}$. 
Hereafter, all random variables and random vectors are defined on this probability space or suitable extensions thereof and have finite second moments.
We write $L^{2} \coloneqq L^{2}(\Omega,\mathcal{F},\mathbb{P})$ to mean the real Hilbert space of square-integrable random variables, identified up to almost-sure equality, equipped with the inner product $\langle a, b \rangle \coloneqq \mathbb{E}\left[ab\right]$. 
Two random variables $a, b \in L^{2}$ are said to be orthogonal if $\langle a, b \rangle = 0$, which we denote by $a \perp b$; likewise, we write $a \perp \mathcal{B}$ if $a \perp b$ for every $b \in \mathcal{B}$, where $\mathcal{B} \subseteq L^{2}$.

Let $\mathcal{B}$ be a closed linear subspace of $L^{2}$.
For every $a\in L^{2}$, $\smash{\Ehat\left[a\cond\mathcal{B}\right]}$ will denote the orthogonal projection of $a$ onto $\mathcal{B}$, i.e., the unique element of $\mathcal{B}$ minimizing the mean-square error.
For linear subspaces $\mathcal{A},\mathcal{B}\subseteq L^{2}$ with $\mathcal{A}\cap\mathcal{B}=\{0\}$, we define their direct sum as $\mathcal{A}\oplus\mathcal{B}\coloneqq\{a+b \colon a\in\mathcal{A},\,b\in\mathcal{B}\}$.

For random vectors $\vect{a}$ and $\vect{b}$, the above definitions apply componentwise, with $\langle \vect{a}, \vect{b} \rangle \coloneqq \mathbb{E}[\vect{a}^\top \vect{b}] = \operatorname{tr}(\mathbb{E}[\vect{a}\vect{b}^\top])$.
\section{PROBLEM FORMULATION}\label{sec:problem formulation}
Consider an $\mathbb{R}^{n}$-valued random state $\vect{x}$ with mean $\est{x}_{0}$ and covariance $\matr{P}_{0}\succ \matr{0}$.
At each time step $t\in\mathbb{N}$ the value of $\mathbf{x}$ is revealed through the scalar signal $s_{t}(\mathbf{x})$, defined by the following time-varying map:
    \begin{equation}\label{eq:signal map}
        s_{t}\colon\mathbb{R}^{n}\to\mathbb{R}, 
        \qquad
        s_{t}(\boldsymbol{\xi}) \coloneqq \vect{c}_{t}^{\top} \boldsymbol{\xi},
    \end{equation}
where $\{\vect{c}_{t} \colon t\in\mathbb{N} \}$ is a deterministic sequence in $\mathbb{R}^{n}\setminus\{\vect{0}\}$.
For $\vect{x}$ to be estimated in every direction, the regressor must keep spanning $\mathbb{R}^{n}$.
We require this to hold uniformly over time by assuming that $\{\vect{c}_{t}\colon t\in\mathbb{N} \}$ is strongly persistently exciting~\citep[Def.~3.4.A]{goodwin2014adaptive}:

\smallskip
\begin{assum}[Strong persistency of excitation]\label{assum:pe}
There\\exist an integer $\bar{t}\geq 0$ and constants
$\underaccent{\bar}{c},\bar{c}>0$ such that
\begin{equation*}
    \underaccent{\bar}{c}\,\matr{I}
    \preceq\sum_{\tau=t}^{t+\bar{t}}\mathbf{c}_{\tau}\mathbf{c}_{\tau}^\top
    \preceq\bar{c}\,\matr{I},
\end{equation*}
for every $t\in\mathbb{N}$. \hfill$\triangle$
\end{assum}

\subsection{Network Structure}
Let $\mathcal{V}\coloneqq\{1,\ldots,r\}$ denote a set of noncooperative agents, each seeking to estimate $\vect{x}$.
The agents are organized as a directed tree rooted at $r$, with edges in the set $\mathcal{E}\subseteq\mathcal{V}\times\mathcal{V}$, where $(j,i)$ means that $j$ is a child of $i$.
While the \emph{root} $r$ has no parent, every other agent has exactly one \emph{parent} and one directed path to $r$: all edges therefore point toward the root.
Agents without \emph{children} are referred to as \emph{leaves} and form the set $\mathcal{L}\subseteq\mathcal{V}$.
The proposed topology well explains how information flows through simplified hierarchies.

\begin{figure}[t]
  \centering
  \input{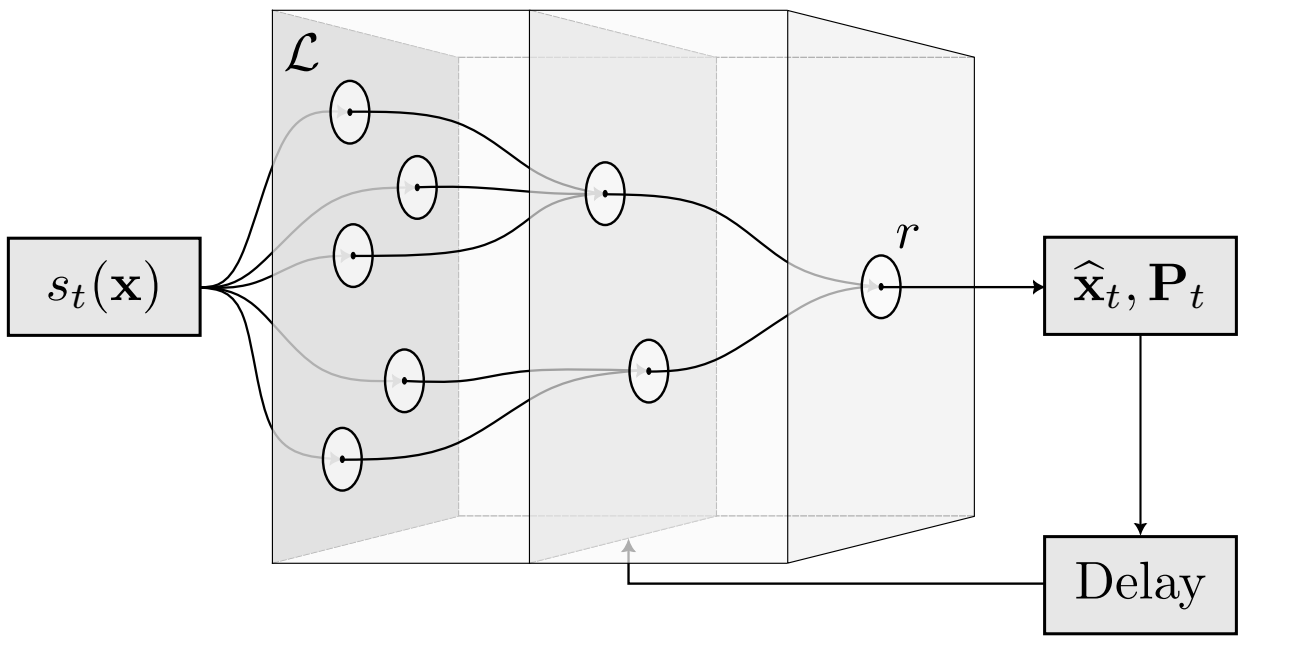}
  \caption{Word-of-Mouth estimation in a directed tree.}
  \label{fig:network}
\end{figure}

The agents are asymmetrically informed via their observations: the leaves observe the signal $s_{t}(\vect{x})$, whereas all other agents simply observe the least-squares signal estimates of their children, with additive noise corrupting both types of observations.
Therefore, the latter learn about the state by \emph{word of mouth}: the farther they sit from the leaves, the more relays their information has gone through. 
As shown in Figure~\ref{fig:network}, both instances admit a common representation in terms of the augmented acyclic graph $(\bar{\mathcal{V}},\bar{\mathcal{E}})$ defined by
    \begin{equation*}
        \bar{\mathcal{V}}\coloneqq\mathcal{V}\cup\{0\},
        \qquad
        \bar{\mathcal{E}}\coloneqq\mathcal{E}\cup\{(0,i)\colon i\in\mathcal{L}\},
    \end{equation*}
where the fictitious agent $0$ represents the signal $s_t(\vect{x})$, and serves as a child to all leaves.
With $\bar{\mathcal{C}}_i\coloneqq\{j\in\bar{\mathcal{V}}\colon(j,i)\in\bar{\mathcal{E}}\}$ denoting the children of agent $i\in\mathcal{V}$, all agents can thus be viewed as observing their children, whether through direct sensing or relayed signal estimates. 
Accordingly, we index observation noise by the edges of $\bar{\mathcal{E}}$: for each $(j,i)\in\bar{\mathcal{E}}$, let
$\smash{\{e_t^{ij}\colon t\in\mathbb{N}\}}$ denote the additive noise sequence corrupting agent $i$'s observations of child $j$.

\smallskip
\begin{assum}[Orthogonality]\label{assum:orth}
For every $(j,i)\in\bar{\mathcal{E}}$, the sequence $\smash{\{e_t^{ij}\colon t\in\mathbb{N}\}}$ has zero mean and constant variance $\varepsilon^{ij}>0$. 
Noise terms are pairwise orthogonal across distinct triples $(j,i,t)$, and each is orthogonal to $\mathbf{x}$. \hfill$\triangle$
\end{assum}

\subsection{Word-of-Mouth Estimation}
\begin{figure}[t]
  \centering
  \begingroup%
\begin{tikzpicture}
\usetikzlibrary{arrows.meta,calc,positioning,fit}%
\definecolor{boxgray}{RGB}{128,128,128}%
\definecolor{panelgray}{RGB}{242,242,242}%
\newcommand{\agentfigurefont}{\normalfont\fontsize{10}{11}\selectfont}%
\newcommand{\agentoutputlabel}{s_t^i(\widehat{\mathbf{x}}_t^i)}%
\newlength{\agentshortarrow}%
\newlength{\agentoutsideinput}%
\newlength{\agentoutsideoutput}%
\newlength{\agentlabelwidth}%
\settowidth{\agentshortarrow}{{\agentfigurefont$\widehat{\mathbf{x}}_t^i$}}%
\settowidth{\agentlabelwidth}{{\agentfigurefont$y_t^i$}}%
\ifdim\agentlabelwidth>\agentshortarrow%
  \setlength{\agentshortarrow}{\agentlabelwidth}%
\fi%
\addtolength{\agentshortarrow}{3mm}%
\settowidth{\agentoutsideinput}{{\agentfigurefont$y_t^{ij}$}}%
\addtolength{\agentoutsideinput}{3mm}%
\settowidth{\agentoutsideoutput}{{\agentfigurefont$\agentoutputlabel$}}%
\addtolength{\agentoutsideoutput}{3mm}%
%
\pgfdeclarelayer{agentwires}%
\pgfdeclarelayer{agentpanel}%
\pgfsetlayers{agentpanel,agentwires,main}%
\tikzset{
  font=\agentfigurefont,line cap=butt,line join=miter,
  box/.style={rectangle,draw=black,line width=0.2mm,
    fill=boxgray,fill opacity=0.188235,draw opacity=1,text opacity=1,
    align=center,inner xsep=2mm,inner ysep=1.5mm,outer sep=0pt},
  arrow/.style={draw=black,line width=0.15mm,
    -{Stealth[length=1.35mm,width=1.1mm]}},
  lab/.style={inner sep=0pt,outer sep=0pt}
}
\node[box,fill=white,fill opacity=0.6] (weight) {Inverse\\variance\\weighting};
\node[box,fill=white,fill opacity=0.6,right=\agentshortarrow of weight] (rls) {RLS};
\node[box,fill=white,fill opacity=0.6,right=\agentshortarrow of rls] (map) {Signal\\model};
\node[fit=(weight)(rls)(map),inner sep=2mm,outer sep=0pt] (panel) {};
\coordinate (prioraxis) at (rls.center |- panel.south);
\node[box,anchor=north] (prior) at ([yshift=-4mm]prioraxis)
  {$\widehat{\mathbf{x}}_{t-1},\mathbf{P}_{t-1}$};

\coordinate (inupper) at ($(weight.west)!0.5!(weight.north west)$);
\coordinate (inlower) at ($(weight.west)!0.5!(weight.south west)$);
\coordinate (entryupper) at (panel.west |- inupper);
\coordinate (entrylower) at (panel.west |- inlower);
\coordinate (startupper) at ([xshift=-\agentoutsideinput]entryupper);
\coordinate (startlower) at ([xshift=-\agentoutsideinput]entrylower);
\coordinate (entrymid) at (panel.west |- weight.center);
\coordinate (startmid) at ([xshift=-\agentoutsideinput]entrymid);
\coordinate (exit) at (panel.east |- map.east);
\coordinate (outputend) at ([xshift=\agentoutsideoutput]exit);

\begin{pgfonlayer}{agentwires}
  \draw[arrow] (startupper) -- (inupper);
  \draw[arrow] (startlower) -- (inlower);
  \draw[arrow] (weight.east) -- (rls.west);
  \draw[arrow] (rls.east) -- (map.west);
  \draw[arrow] (map.east) -- (outputend);
  \draw[arrow] (prior.north) -- (rls.south);
\end{pgfonlayer}
\begin{pgfonlayer}{agentpanel}
  \path[fill=panelgray,fill opacity=0.8,draw=black,draw opacity=1,line width=0.075mm,densely dotted,line cap=round]
    (panel.south west) rectangle (panel.north east);
\end{pgfonlayer}

\node[lab] at ($(startmid)!0.5!(entrymid)$) {$y_t^{ij}$};
\node[lab,above=1.2mm] at ($(exit)!0.5!(outputend)$) {$\agentoutputlabel$};
\node[lab,above=1.2mm] at ($(weight.east)!0.5!(rls.west)$) {$y_t^i$};
\node[lab,above=1.2mm] at ($(rls.east)!0.5!(map.west)$) {$\widehat{\mathbf{x}}_t^i$};
\end{tikzpicture}%
\endgroup%
  \caption{The Word-of-Mouth estimation pipeline.}
  \label{fig:agent}
\end{figure}
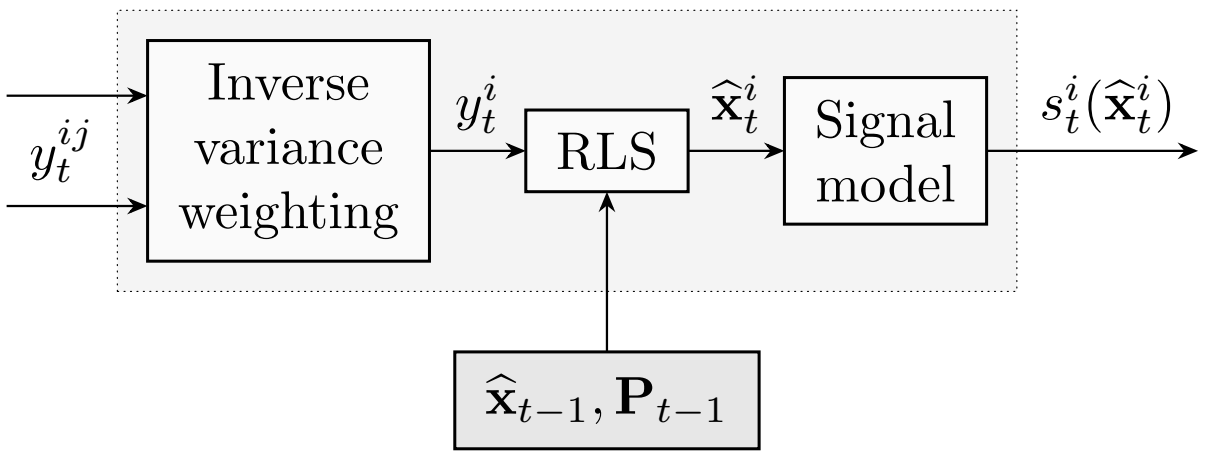

Each agent estimates the state from common prior information and local observations.
At time step $t\in\mathbb{N}$, all local updates start from $(\est{x}_{t-1}, \matr{P}_{t-1})$, with initial pair $(\est{x}_{0}, \matr{P}_{0})$.
Updates proceed from the leaves to the root, following the pipeline in Figure~\ref{fig:agent}: local observations are aggregated and used in an RLS update, after which each agent except the root transmits its signal estimate to its parent.

To accommodate noisy and possibly indirect observations, each $i\in\mathcal{V}$ maintains a local time-varying model $s_{t}^{i}\colon\mathbb{R}^{n}\to\mathbb{R}$ of the signal~\eqref{eq:signal map}, whose form we specify below.
For each $j\in\bar{\mathcal{C}}_{i}$, the observation process $\smash{\{y_{t}^{ij}\colon t\in\mathbb{N}\}}$ is defined by
    \begin{equation}\label{eq:obs at edge true}
        y_{t}^{ij} 
        = 
        \begin{cases}
            s_{t}(\vect{x}) + e_{t}^{i0}, \quad&\textrm{if }i\in\mathcal{L}, \\
            s_{t}^{j}(\est{x}_{t}^{j}) + e_{t}^{ij}, \quad&\textrm{otherwise},
        \end{cases}
    \end{equation}
where $\est{x}_{t}^{j}$ is the RLS state estimate computed by one child $j$ at time $t$, and $\smash{s_{t}^{j}(\est{x}_{t}^{j})}$ is its corresponding signal estimate.
This construction rests on the following ansatz: the RLS update is affine in the common estimate $\est{x}_{t-1}$ and incoming observations $\smash{\{y_{t}^{ij}\colon j\in\bar{\mathcal{C}}_{i}\}}$, suggesting that the local signal models might be chosen to preserve this structure through successive relays.
We therefore seek a representation separating the current signal, the contribution of the common estimate, and accumulated noise:
    \begin{equation}\label{eq:obs at edge ansatz}
        y_{t}^{ij} 
        = 
        a_{t}^{ij} ( s_{t}(\mathbf{x}) + w_{t}^{ij} ) + b_{t}^{ij},
        \qquad
        (j,i)\in\bar{\mathcal{E}}.
    \end{equation}
Here $\{w_{t}^{ij}\colon t\in\mathbb{N}\}$ is the \emph{effective} noise process incorporating all the upstream observation noise and the disturbance on the receiving edge, with zero mean and variance parameter $\smash{\varphi_{t}^{ij}}$.
The validity of representation \eqref{eq:obs at edge ansatz} and its variance interpretation are established in Section~\ref{sec:structural results}.

Subject to~\eqref{eq:obs at edge ansatz}, the normalized observations $\smash{(y_t^{ij}-b_t^{ij})/a_t^{ij}}$ differ from one another only in their effective noise terms, with smaller $\smash{\varphi_t^{ij}}$ giving less uncertain information on $s_t(\vect{x})$.
For each $i\in\mathcal{V}$, we define the aggregate observation process $\{y_t^i\colon t\in\mathbb{N}\}$ using inverse-variance weighting~\citep[Thm.~7A.1]{jazwinski1970stochastic}:
    \begin{equation}\label{eq:aggregation}
        y_{t}^{i}
        \coloneqq
        ( \varphi_{t}^{i} )^{\eta}
        \sum_{j\in\bar{\mathcal{C}}_{i}} \dfrac{y_{t}^{ij} - b_{t}^{ij}}{a_{t}^{ij} \varphi_{t}^{ij}},
        \qquad
        \frac{1}{\varphi_{t}^{i}}
        \coloneqq\sum_{j\in\bar{\mathcal{C}}_i} \dfrac{1}{\varphi_{t}^{ij}},
    \end{equation}
where $\eta\in[0,1]$ is a constant normalization parameter. 
Thus, all incoming observations contribute, with each normalized observation weighted in proportion to its \emph{precision} $1/\varphi_t^{ij}$.
Substituting~\eqref{eq:obs at edge ansatz} yields
    \begin{equation}\label{eq:obs at node ansatz}
        y_{t}^{i}
        =
        a_{t}^{i} \, (s_{t}(\vect{x}) + w_{t}^{i}),
        \qquad
        w_{t}^{i}
        \coloneqq\varphi_{t}^{i} \sum_{j\in\bar{\mathcal{C}}_{i}} \dfrac{w_{t}^{ij}}{\varphi_{t}^{ij}},
    \end{equation}
where $a_t^i\coloneqq(\varphi_t^i)^{\eta-1}$ and $s_t^i\coloneqq a_t^i s_t$ is the signal model of $i$.
Each agent combines the common pair $(\est{x}_{t-1},\matr{P}_{t-1})$ with the normalized aggregate $y_t^i/a_t^i$ through the standard RLS update \citep[Sec.~7.3, Ex.~7.2]{jazwinski1970stochastic}:
    \begin{subequations}\label{eq:RLS update}
    \begin{align}
        \est{x}_{t}^{i} 
        & = \est{x}_{t-1} + \dfrac{ \matr{P}_{t-1} \vect{c}_{t} }{ \vect{c}_{t}^\top \matr{P}_{t-1} \vect{c}_{t} + \varphi_{t}^{i}  } \left( \dfrac{y_{t}^{i}}{a_{t}^{i}} - s_{t}(\est{x}_{t-1}) \right), \label{eq:RLS update state estimate} \\
        \matr{P}_{t}^{i} 
        & = \matr{P}_{t-1} - \dfrac{ \matr{P}_{t-1} \vect{c}_{t} \vect{c}_{t}^\top \matr{P}_{t-1} }{ \vect{c}_{t}^\top \matr{P}_{t-1} \vect{c}_{t} + \varphi_{t}^{i}  }.
        \label{eq:RLS update variance}
    \end{align}
    \end{subequations}
Each $i\in\mathcal{V}\setminus\{r\}$ transmits $s_t^i(\est{x}_t^i)$ to its parent, providing a noisy observation at the next level, with $\eta$ controlling the signal scale relative to the added noise.
The root $r$ closes the round by setting $(\est{x}_t,\matr{P}_t)\coloneqq(\est{x}_t^r,\matr{P}_t^r)$ and broadcasting this pair.
Before the next time step, each agent discards its private history of measurements and estimates and adopts the broadcast pair unconditionally.
Authority thus enters through the common prior information: past observations contribute to subsequent updates only through the root’s summary, which supersedes every local assessment.

\subsection{Problem Statement}
The root broadcasts its estimate and error covariance to all agents, so mean-square estimation at the root determines learning throughout the network.
The agents are nevertheless noncooperative and do not relay strategic information to their parents.
A first question is whether the prescribed protocol minimizes each agent's mean-square error within its linear information space.
Notably, however, local optimality need not imply network optimality~\citep{witsenhausen1968counterexample}, a distinction that motivates joint design of network topology and estimation protocols.
We further ask whether $\{\matr{P}_t\colon t\in\mathbb{N}\cup\{0\}\}$ converges, whether its limit is zero and, if so, at what rate it converges.
We address these questions by reducing the network to an equivalent single-agent RLS update and by deriving order-matching upper and lower bounds on $\matr{P}_t$.
Our bounds relate the rate of convergence to the structure of the tree and suggest that learning within hierarchies is a slow process in general.

\section{STRUCTURAL RESULTS}\label{sec:structural results}
We first establish representation~\eqref{eq:obs at edge ansatz} and make precise the sense in which each local update is optimal. 
The network thereby reduces to a single agent, whose observation noise accounts for the entire hierarchy.

\subsection{LMMSE Formulation}
Optimality is meaningful only relative to what each agent knows, i.e., the information that the protocol retains. 
Let $\mathcal{F}_0^i = \mathcal{F}_0 \coloneqq \operatorname{span}(\{\mathbf{1}\})$ and, for $t \in \mathbb{N}$,
    \begin{align*}
    \mathcal{F}_{t}^{i}
    & \coloneqq \mathcal{F}_{t-1} + \operatorname{span}(\{y_{t}^{ij} \colon j\in\bar{\mathcal{C}}_{i}\}), \\
    \mathcal{F}_{t}
    & \coloneqq \mathcal{F}_{t-1} + \operatorname{span}(\{\est{x}_{t}\}),
    \end{align*}
where the span of a vector is that of its components; being finite-dimensional, these subspaces are closed. 
Accordingly, $\mathcal{F}_t$ records the broadcast history, to which $\mathcal{F}_t^i$ adds only what agent~$i$ observes at time~$t$, since local
histories are discarded after every round. 
With $q_t\coloneqq\vect{c}_t^\top\matr{P}_{t-1}\vect{c}_t$, the representation~\eqref{eq:obs at edge ansatz} holds as follows.

\smallskip
\begin{prop}
\label{prop:observation model at edges}
    Let Assumption~\ref{assum:orth} hold. 
    Then \eqref{eq:obs at edge ansatz} holds for every $t\in\mathbb{N}$ with the following properties:
    \begin{itemize}
        \item[(i)] 
        For every $(j,i)\in\mathcal{E}$, the parameters are constructed recursively as
        \begin{align*}
            a_t^{ij}
            &\coloneqq \frac{a_t^j q_t}{q_t+\varphi_t^j},
            &
            b_t^{ij}
            &\coloneqq  \frac{a_t^j\varphi_t^j}{q_t+\varphi_t^j}\, s_t(\widehat{\mathbf{x}}_{t-1}),
            \\
            w_t^{ij}
            &\coloneqq  w_t^j+\frac{e_t^{ij}}{a_t^{ij}},
            &
            \varphi_t^{ij}
            &\coloneqq  \varphi_t^j+\frac{\varepsilon^{ij}}{(a_t^{ij})^2},
        \end{align*}
        with $a_{t}^{i0} \coloneqq 1, b_{t}^{i0} \coloneqq 0, w_{t}^{i0} \coloneqq e_{t}^{i0}, \varphi_{t}^{i0} \coloneqq \varepsilon^{i0}$ for $i\in\mathcal{L}$.
        \item[(ii)] For every $(j,i)\in\bar{\mathcal{E}}$, ${a_t^{ij}}$ and $\smash{\varphi_t^{ij}}$ are strictly positive, deterministic, and computable offline, and $\smash{b_t^{ij}\in\mathcal{F}_{t-1}}$.
        \item[(iii)] For every $(j,i)\in\bar{\mathcal E}$ and $t\in\mathbb N$,  ${w_t^{ij}}$ is zero-mean with variance $\smash{\varphi_t^{ij}}$.
        Moreover, for a fixed $i\in\mathcal{V}$, the variables $\smash{w_t^{ij}}$ are pairwise orthogonal across distinct pairs $(j,t)$, and each is orthogonal to $\mathcal F_{t-1}$ and to $\mathbf{x}$.
        \hfill$\triangle$
    \end{itemize}
\end{prop}

\begin{pf}
    See Appendix.\hfill$\square$
\end{pf}

By Proposition~\ref{prop:observation model at edges}(iii), effective noise terms reaching agent~$i$ are orthogonal, whence $\varphi_t^i$ in~\eqref{eq:aggregation} is precisely the variance of the aggregate noise $w_t^i$ in~\eqref{eq:obs at node ansatz}.
This variance depends on the common prior information only through~$q_t$, and it does so monotonically: a larger $q_t$ raises every relay gain $\smash{a_t^{ij}}$, so that the noise on each receiving edge then weighs less after normalization. 
As a result, $\varphi_t^i$ is constant at each leaf and strictly decreasing in~$q_t$ otherwise; for a fixed regressor, it is hence nonincreasing in $\matr{P}_{t-1}$ in the L\"owner order. This
compares two priors at the same time step, and entails no monotonicity over time. The
same noise statistics render each local update optimal.

\smallskip
\begin{prop}
\label{prop:optimality}
    Let Assumption~\ref{assum:orth} hold. 
    For each $i\in\mathcal{V}$ and $t\in\mathbb{N}$, $\mathcal{F}_{t}^{i}$ admits the orthogonal decomposition
        \begin{equation*}
            \mathcal{F}_t^i
            = 
            \mathcal{F}_{t-1} \oplus \operatorname{span}\left(\left\{ \frac{y_t^{ij}-b_t^{ij}}{a_t^{ij}} -s_t(\widehat{\mathbf{x}}_{t-1}) \colon j\in\bar{\mathcal{C}}_i \right\}\right),
        \end{equation*}
    and~\eqref{eq:RLS update state estimate}--\eqref{eq:RLS update variance} satisfy
        \begin{align*}
            \widehat{\mathbf{x}}_t^i
            &= \Ehat[\mathbf{x}\cond\mathcal{F}_t^i],
            \\
            \matr{P}_t^i
            &= \mathbb{E}\bigl[  (\mathbf{x}-\widehat{\mathbf{x}}_t^i) (\mathbf{x}-\widehat{\mathbf{x}}_t^i)^\top \bigr],
        \end{align*}
    where $\smash{\matr{P}_t^i\succ \matr{0}}$ is deterministic and computable offline. \hfill$\triangle$
\end{prop}

\begin{pf}
    See Appendix.\hfill$\square$
\end{pf}

Proposition~2 gives (6a)--(6b) its statistical meaning: each update returns the unbiased LMMSE estimate of~$\vect{x}$ given $\mathcal{F}_t^i$ and its error
covariance; in particular, $q_t$ is the variance of $s_t(\vect{x})-s_t(\est{x}_{t-1})$. 
In addition,~\eqref{eq:aggregation} is a lossless aggregation rule.
Even though $\mathcal{F}_t^i$ contains every incoming observation, the single aggregate observation $y_t^i$ suffices, since inverse-variance weights are optimal, in the Gauss--Markov sense, for orthogonal noises~\citep[Thm.~7A.1]{jazwinski1970stochastic}. 
These claims are, however, relative to the prescribed information: no linear estimator fares better with what the protocol lets an agent know, yet nothing here suggests that discarding local histories, or deferring to the root, is itself optimal.
 
Learning, should it occur, is self-defeating. 
As $t\to\infty$, if $\matr{P}_t\to{\matr{0}}$, then $q_t\to0$ by Assumption~\ref{assum:pe} and, following from Proposition~\ref{prop:observation model at edges}(i), $\smash{a_t^{ij}\to0}$ on every $(j,i)\in\mathcal{E}$, while $\varphi_t^i\to\infty$ for every $i\in\mathcal{V}\setminus\mathcal{L}$.
This mechanism parallels the feedback effect in~\cite{vives1993fast}, where improved public information weakens the agents' responses to their private signals and thereby slows the revelation of further information.
Each estimate remains optimal for the agent that produces it, yet becomes less informative to the agent that observes it.

\smallskip
\begin{rem}[Electrical networks]
\label{re:electrical networks}
    The recursion in Proposition~\ref{prop:observation model at edges}(i) admits an electrical
    interpretation~\citep{doyle1984random,barooah2008estimation}.
    Across a relay, noise variances add as \emph{resistances} in series; at aggregation, precisions add as \emph{conductances} in parallel. Hence $\varphi_t^i$ is the effective resistance between the source~$0$ and agent~$i$ in the subtree rooted at~$i$.
    Long chains of relays thus accumulate resistance, while additional branches provide conductance; as $q_t$ decreases, the relay resistances grow.
    \hfill$\triangle$
\end{rem}

\subsection{Single-Agent RLS Formulation}
At the root,~\eqref{eq:obs at node ansatz} gives $y_t^r/a_t^r=s_t(\vect{x})+w_t^r$. 
By Proposition~\ref{prop:observation model at edges}, $w_t^r$ has zero mean and variance $\varphi_t^r$, and is orthogonal to~$\vect{x}$ and across time.
The broadcast pair is therefore precisely a single-agent RLS update, had that agent observed the signal directly in additive white noise of variance~$\varphi_t^r$.
In information form,~\eqref{eq:RLS update variance} at the root reads as
    \begin{equation}
    \label{eq:RLS info form single agent}
        \matr{P}_{t} 
        = \biggl(\matr{P}_{t-1}^{-1} + \dfrac{\vect{c}_{t} \vect{c}_{t}^{\top}}{\varphi_{t}^{r}}\biggr)^{-1},
    \end{equation}
i.e., each round adds the rank-one precision ${\vect{c}_t\vect{c}_t^\top/\varphi_t^r\succeq\matr{0}}$. 
Consequently, $\{\matr{P}_t\colon t\in\mathbb{N}\cup\{0\}\}$ is nonincreasing in the L\"owner order and bounded below, hence convergent.
\section{CONVERGENCE ANALYSIS}\label{sec:convergence analysis}
In this section, we show that the network learns, that is, $\matr{P}_t\to \matr{0}$ as $t\to\infty$, albeit slowly: our bounds on $\matr{P}_t$ are of matching order in~$t$, and their exponent depends only on~$\eta$ and on how far the root sits from the leaves.

For each $i\in\mathcal{V}$, let $\delta_i$ be the length of the shortest directed path in $(\mathcal{V},\mathcal{E})$ connecting a leaf to~$i$, i.e., the least number of relays between agent~$i$ and the signal $s_{t}(\vect{x})$.
Thus, $\delta_i=0$ if $i\in\mathcal{L}$, and $\smash{\delta_i=1+\min_{j\in\bar{\mathcal{C}}_i}\delta_j}$ otherwise.
We first relate the variance $\smash{\varphi^i_t}$ of the aggregate noise in~\eqref{eq:obs at node ansatz} to~$q_t$.

\smallskip
\begin{lem}
\label{lem:order of contact}
    Let Assumptions~\ref{assum:pe} and~\ref{assum:orth} hold, and define
        \begin{equation*}
            \alpha_{i} 
            \coloneqq
            \dfrac{2(4-2\eta)^{\delta_{i}}-2}{3 - 2\eta}
        \end{equation*}
    for every $i\in\mathcal{V}$.
    There exist constants $\underaccent{\bar}{\varphi}, \bar{\varphi}>0$ such that
        \begin{equation}\label{eq:bound on effective noise variance}
            \underaccent{\bar}{\varphi} \, q_{t}^{-\alpha_{i}} \leq \varphi_{t}^{i} \leq \bar{\varphi} \, q_{t}^{-\alpha_{i}}
        \end{equation}
    for every  $i\in\mathcal{V}$ and $t\in\mathbb{N}$.
    \hfill$\triangle$
\end{lem}

\begin{pf}
    See Appendix.\hfill$\square$
\end{pf}

Lemma~\ref{lem:order of contact} quantifies the feedback effect of Section~\ref{sec:structural results}: should $q_t$ vanish, $\varphi^i_t$ diverges for every $i\in\mathcal{V}\setminus\mathcal{L}$, with an exponent that depends on $(\mathcal{V},\mathcal{E})$ only through~$\delta_i$.
As its proof shows, $\smash{\varphi^{ij}_t}$ obeys bounds of the form~\eqref{eq:bound on effective noise variance} on every $(j,i)\in\mathcal{E}$, with exponent $\alpha^{ij}\coloneqq(4-2\eta)\,\alpha_j+2>\alpha_j$; in words, every relay strictly increases the exponent.
Because $q_t$ is bounded, the sum of precisions in~\eqref{eq:aggregation} is of the same order as its term with the smallest exponent, whence $\smash{\alpha_i=\min_{j\in\bar{\mathcal{C}}_i}\alpha^{ij}}$, attained exactly by the children~$j$ with $\delta_j=\delta_i-1$.
All other children are ignored asymptotically: the normalized observation of child~$j$ enters $\smash{y^i_t/a^i_t}$ with weight $\smash{\varphi^i_t/\varphi^{ij}_t}$, of order $\smash{q_t^{\alpha^{ij}-\alpha_i}}$, which vanishes with $q_t$ unless $\delta_j=\delta_i-1$.

\smallskip
\begin{rem}[Large deviations]
    This path selection has a large-deviations interpretation: as $\nu\to 0$, a finite sum of terms $\exp(-\alpha/\nu)$ is dominated, on a logarithmic scale, by the term with the smallest~$\alpha$ \citep[Lem.~1.2.15]{dembo2009large}.
    Indeed, with $\nu_t\coloneqq1/\log(1/q_t)$ for $q_t<1$, we have $\smash{q_t^{\alpha^{ij}}=\exp(-\alpha^{ij}/\nu_t})$, so, whenever $q_t\to0$, each precision in~\eqref{eq:aggregation} satisfies $\smash{\nu_t\log(1/\varphi^{ij}_t)\to-\alpha^{ij}}$, and their sum fulfills $\nu_t\log(1/\varphi^{i}_t)\to-\alpha_i$.
    Much as rare events occur in the least unlikely way, agent~$i$ learns asymptotically only through shortest directed paths from the leaves.
\hfill$\triangle$
\end{rem}

Inserted into~\eqref{eq:RLS info form single agent}, Lemma~\ref{lem:order of contact} bounds the precision $\mathbf{c}_t\mathbf{c}_t^\top/\varphi^r_t$ added in each
round between multiples of $\smash{q_t^{\alpha_r}\mathbf{c}_t\mathbf{c}_t^\top}$.
Unless $\delta_r=0$, this precision thus vanishes with~$q_t$: the more the
network has learned, the less each round adds. Our main result shows
that $\mathbf{P}_t$ converges to zero, nevertheless.

\smallskip
\begin{thm}
\label{thm:rate of convergence}
    Let Assumptions~\ref{assum:pe} and~\ref{assum:orth} hold, and define
        \begin{equation*}
            \beta
            \coloneqq
            \dfrac{1}{1 + \alpha_{r}}
            =
            \dfrac{3 - 2\eta}{2(4 - 2\eta)^{\delta_{r}} + 1 - 2\eta}.
        \end{equation*}
    There exist constants $\underaccent{\bar}{\pi}, \bar{\pi}>0$ such that
        \begin{equation}
        \label{eq:bound on error covariance}
            \underaccent{\bar}{\pi}\,(t+1)^{-\beta} \, \matr{I}
            \preceq
            \matr{P}_{t-1}
            \preceq
            \bar{\pi} \, (t+1)^{-\beta} \, \matr{I}
        \end{equation}
    for every $t\in\mathbb{N}$.
    \hfill$\triangle$
\end{thm}

\begin{pf}
    See Appendix.\hfill$\square$
\end{pf}
Theorem~\ref{thm:rate of convergence} answers the questions raised in Section~\ref{sec:problem formulation}. 
By Proposition~\ref{prop:optimality} at the root, the mean-square error of $\est{x}_t$ is $\langle \vect{x} - \est{x}_t, \, \vect{x} - \est{x}_t \rangle = \operatorname{tr}(\matr{P}_t)$. 
Using the upper bound in~\eqref{eq:bound on error covariance}, we obtain $\operatorname{tr}(\matr{P}_t) \leq n\bar{\pi}\,(t+2)^{-\beta}$, so that $\est{x}_t$ converges to $\vect{x}$ in mean square. 
Since $\matr{P}_t^i \preceq \matr{P}_{t-1}$ by~\eqref{eq:RLS update variance}, the same holds for the estimate $\est{x}_t^i$ of every $i \in \mathcal{V}$: hierarchical networks do learn. 
At the same time, by the lower bound in~\eqref{eq:bound on error covariance}, this rate cannot be improved in any direction of $\mathbb{R}^n$.
 
The exponent $\beta$ itself reflects how far the root sits from the leaves.
If $\delta_r=0$, that is, $\mathcal{V}=\{r\}$, then $\beta=1$, which is the rate of a single agent observing the signal.
Otherwise, $\smash{\alpha_r=2\sum_{k=0}^{\delta_r-1}(4-2\eta)^k}\ge 2$, so that $\beta\le 1/3$, with equality if $\delta_r=1$, and $\beta$ is at least halved by every additional relay on a shortest directed path from a leaf to the root.
No other feature of $(\mathcal{V},\mathcal{E})$ enters~$\beta$: neither the number of agents, nor the number of children per agent, nor the length of any other directed path.
By contrast, $\beta$ depends on~$\eta$ as soon as $\delta_r\ge2$, and is then strictly increasing in~$\eta$.
Both of these effects are illustrated in Table~\ref{tab:parameter values}.
For $\eta=1$, the aggregation rule~\eqref{eq:aggregation} normalizes the incoming observations, in that the weights $\smash{(\varphi^i_t)^\eta/\varphi^{ij}_t}$ it assigns to the normalized observations sum to one, and $\beta=1/(2^{\delta_r+1}-1)$, which, for a directed path with scalar~$\vect{x}$ and $\vect{c}_t=1$, is the rate established by~\citet{krishnamurthy2024slow}.

\newlength{\tablecolwidth}
\begin{table}[tb]
    \centering
    \caption{Values of $\beta$ for different $\eta$ and $\delta_r$.}
    \label{tab:parameter values}
    \renewcommand{\arraystretch}{1.4}
    \setlength{\tabcolsep}{8pt}
    \settowidth{\tablecolwidth}{1/43}
    \begin{tabular}{ccccc}
        \Hline
                      & \multicolumn{4}{c}{$\delta_r$} \\
        \cline{2-5}
        \makebox[\tablecolwidth]{$\eta$}
        & \makebox[\tablecolwidth]{$0$}
        & \makebox[\tablecolwidth]{$1$}
        & \makebox[\tablecolwidth]{$2$}
        & \makebox[\tablecolwidth]{$3$} \\
        \hline
        $0$           & 1 & 1/3 & 1/11 & 1/43 \\
        $1/2$         & 1 & 1/3 & 1/9  & 1/27 \\
        $1$           & 1 & 1/3 & 1/7  & 1/15 \\
        \Hline
    \end{tabular}
\end{table}
\section{Numerical Examples}\label{sec:numerical examples}
We evaluate convergence of the sequence generated by~\eqref{eq:RLS info form single agent} for different values of the pair\footnote{Code available at \href{https://github.com/andreadacol98/IFAC_SYSID_2027} {\texttt{github.com/andreadacol98/IFAC\_SYSID\_2027}}.}~$(\eta, \delta_{r})$.
Let $n=2$, and consider $\vect{x} = [x_1, x_2]^\top$ with covariance matrix $\matr{P}_{0} = \operatorname{diag}(2,1)$.
The agents seek to estimate the unknown coefficients of
\begin{equation*}
    s_{t}(x_1,x_2) = x_{1} \cos(\omega t) + x_{2} \rho \sin(\omega t),
\end{equation*}
where $\omega,\rho > 0$ are known parameters.
Whenever $\omega\notin\pi\mathbb{N}$, the associated regressor $\vect{c}_t=[\cos(\omega t),\,\rho\sin(\omega t)]^\top$ satisfies Assumption~\ref{assum:pe} with\footnote{Under the same restriction on $\omega$, suitable constants
$\smash{\underaccent{\bar}{c},\bar{c}>0}$ can be obtained for every $\smash{\bar{t}\in\mathbb{N}}$ using the geometric-sum identity underlying the Dirichlet kernel~\citep[Sec.~8.13]{rudin1976principles}.} $\bar{t}=1$ and
\begin{align*}
    \underaccent{\bar}{c} & =\min\{1,\rho^2\}\bigl(1-|\cos(\omega)|\bigr), \\
    \bar{c} & =\max\{1,\rho^2\}\bigl(1+|\cos(\omega)|\bigr).
\end{align*}
Multiples of $\pi$ are excluded because $\sin(\omega t)=0$ at every $t\in\mathbb{N}$, so the second component of the state vector cannot be observed.
Hereafter we set $\omega=1$ and $\rho=2$, which yield $\underaccent{\bar}{c}=1-\cos(1)\approx0.46$ and $\bar{c}=4+4\cos(1)\approx6.16$.
Every observation-noise sequence has unit variance, $\smash{\varepsilon^{ij}=1}$ for all $(j,i)\in\bar{\mathcal{E}}$, and the sequences satisfy Assumption~\ref{assum:orth}.

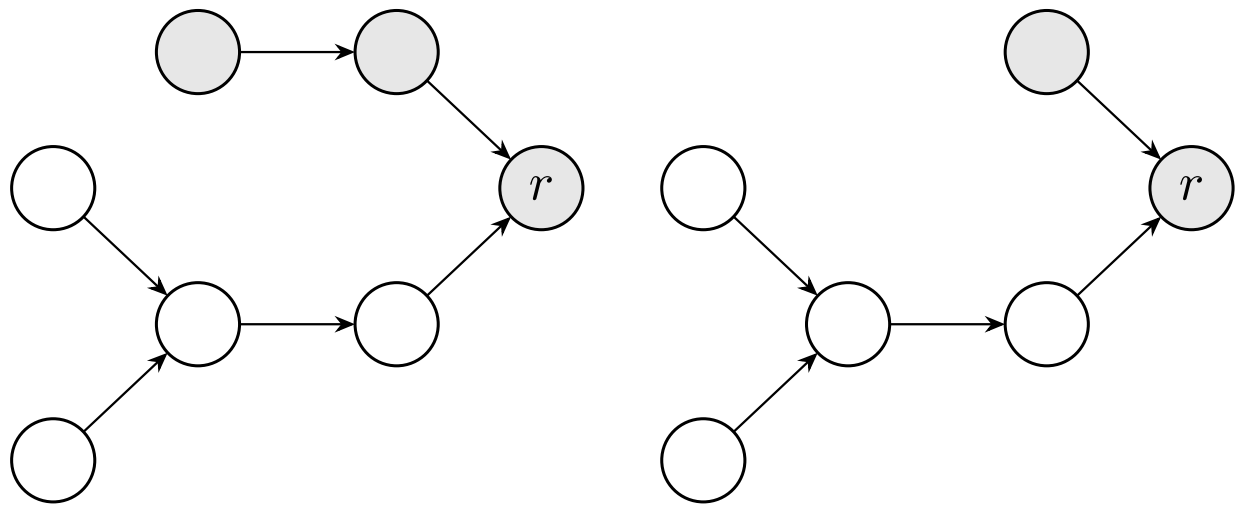
\begin{figure}[t]
    \centering
    \subfloat[$\delta_{r} = 2$. \label{fig:tree-a}]{%
        \begin{tikzpicture}[
    x=1mm, y=1mm,
    agent/.style={circle, draw={rgb,255:red,0;green,0;blue,0}, line width=0.2mm,
        minimum size=\diam mm, inner sep=0pt, outer sep=0pt},
    onpath/.style={agent, fill={rgb,255:red,128;green,128;blue,128}, fill opacity=0.188235, text opacity=1},
    edge/.style={line width=0.15mm, -{Stealth[length=1.35mm, width=1.1mm]}},
    every node/.append style={font=\normalfont\fontsize{10}{11}\selectfont}]
  \def\Wfig{38}     
  \def\diam{5.5}    
  \def\dy{9}        
  \pgfmathsetmacro{\mrg}{\diam/2 + 0.105}
  \pgfmathsetmacro{\W}{\Wfig - 2*\mrg}   
  \pgfmathsetmacro{\dx}{(2*\W - sqrt(\W*\W + 3*\dy*\dy))/3}
  \path[use as bounding box] (-\mrg, -2*\dy-\mrg) rectangle (\W+\mrg, \dy+\mrg);
  \node[onpath] (r)  at (\W,     0)      {$r$};
  \node[onpath] (a2) at (\W-\dx, \dy)    {};
  \node[onpath] (l1) at (\dx,    \dy)    {};   
  \node[agent]  (a6) at (\W-\dx, -\dy)   {};
  \node[agent]  (a5) at (\dx,    -\dy)   {};
  \node[agent]  (a3) at (0,      0)      {};   
  \node[agent]  (a4) at (0,      -2*\dy) {};   
  \draw[edge] (l1) -- (a2);
  \draw[edge] (a2) -- (r);
  \draw[edge] (a6) -- (r);
  \draw[edge] (a5) -- (a6);
  \draw[edge] (a3) -- (a5);
  \draw[edge] (a4) -- (a5);
\end{tikzpicture}}%
    \hspace{5mm}%
    \subfloat[$\delta_{r} = 1$. \label{fig:tree-b}]{%
        \begin{tikzpicture}[
    x=1mm, y=1mm,
    agent/.style={circle, draw={rgb,255:red,0;green,0;blue,0}, line width=0.2mm,
        minimum size=\diam mm, inner sep=0pt, outer sep=0pt},
    onpath/.style={agent, fill={rgb,255:red,128;green,128;blue,128}, fill opacity=0.188235, text opacity=1},
    edge/.style={line width=0.15mm, -{Stealth[length=1.35mm, width=1.1mm]}},
    every node/.append style={font=\normalfont\fontsize{10}{11}\selectfont}]
  \def\Wfig{38}     
  \def\diam{5.5}    
  \def\dy{9}        
  \pgfmathsetmacro{\mrg}{\diam/2 + 0.105}
  \pgfmathsetmacro{\W}{\Wfig - 2*\mrg}   
  \pgfmathsetmacro{\dx}{(2*\W - sqrt(\W*\W + 3*\dy*\dy))/3}
  \path[use as bounding box] (-\mrg, -2*\dy-\mrg) rectangle (\W+\mrg, \dy+\mrg);
  \node[onpath] (r)  at (\W,     0)      {$r$};
  \node[onpath] (l1) at (\W-\dx, \dy)    {};   
  \node[agent]  (a5) at (\W-\dx, -\dy)   {};
  \node[agent]  (a4) at (\dx,    -\dy)   {};
  \node[agent]  (a2) at (0,      0)      {};   
  \node[agent]  (a3) at (0,      -2*\dy) {};   
  \draw[edge] (l1) -- (r);
  \draw[edge] (a5) -- (r);
  \draw[edge] (a4) -- (a5);
  \draw[edge] (a2) -- (a4);
  \draw[edge] (a3) -- (a4);
\end{tikzpicture}}
    \caption{
    Directed trees used in the numerical examples.
    The shaded agents form the shortest directed path from a leaf to the root.
    The fictitious agent~$0$ and its outgoing edges are omitted.
    }
    \label{fig:trees}
\end{figure}

We run the covariance recursion for $10^5$ time steps, considering $\eta\in\{0,1\}$ and the two directed trees in
Figure~\ref{fig:trees}.
The resulting convergence rates are compared with the exponents in Table~\ref{tab:parameter values}.
Since $\matr{P}_t$ is deterministic and can be computed offline, these experiments require no realizations of the state or observation noise.

\begin{figure*}[t]
    \centering
    \includegraphics{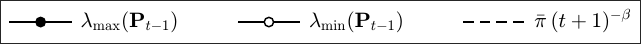}\par\nointerlineskip
    \vspace{0.1mm - 1.5pt}
        \subfloat[$\delta_{r} = 2, \, \eta = 0, \, \beta = 1/11$. \label{fig:plot tree-a unnormlized}]{%
        \includegraphics{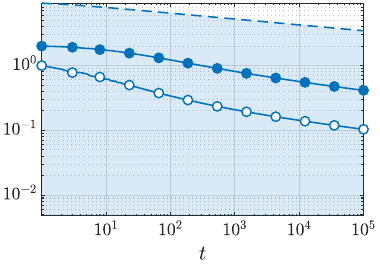}}%
    \hfill
        \subfloat[$\delta_{r} = 2, \, \eta = 1, \beta = 1/7$. \label{fig:plot tree-a normalized copy}]{%
        \includegraphics{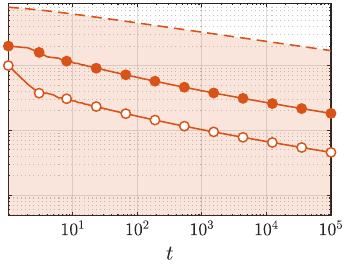}}%
    \hfill
        \subfloat[$\delta_{r} = 1, \eta = 1, \beta = 1/3$. \label{fig:plot tree-b normalized}]{%
        \includegraphics{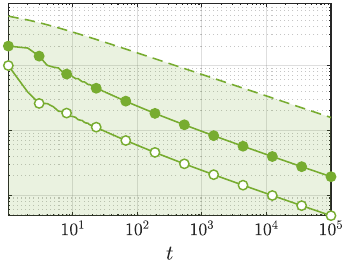}}
    \caption{
        Eigenvalues of $\mathbf{P}_{t-1}$ on logarithmic axes common to all three panels. 
        Panels~(a) and~(b) correspond to the tree in Figure~\ref{fig:trees}(a), and panel~(c) to that in Figure~\ref{fig:trees}(b). 
        The shaded region lies below the upper bound in~\eqref{eq:bound on error covariance}; the lower bound is omitted, as it lies many orders of magnitude below the smallest eigenvalue.
    }
  \label{fig:plots}
\end{figure*}

\subsection{Results}
Figure~\ref{fig:plots} shows the time trajectories of the eigenvalues of $\mathbf{P}_{t-1}$ and the upper bound in~\eqref{eq:bound on error covariance} on logarithmic axes.
In these coordinates, the two bounds in~\eqref{eq:bound on error covariance} are vertical translates of each other, with common slope ${-\beta t/(t+1)}$, which converges to ${-\beta}$ as ${t\to\infty}$.
The eigenvalues lie between the two bounds at all times, so that ${\log\lambda(\mathbf{P}_{t-1})/\log(t+1)}\to{-\beta}$ as well.
Learning is therefore slow: for large $t$, reducing the eigenvalues by one decade requires extending the time horizon by a factor of ${10^{1/\beta}}$, that is, $10^{11}$, $10^{7}$, and $10^{3}$ in panels~(a),~(b), and~(c), respectively.

At $t=10^5$, the bounds are already close to their asymptotic slope, while the eigenvalue trajectories remain visibly steeper.
For the eigenvalues, however, slopes between consecutive time steps continue to oscillate as the regressor changes direction.
We therefore interpret their slopes as changes measured over time windows of asymptotically fixed width on the logarithmic time axis, namely $[t,\lfloor kt\rfloor]$ for a fixed $k>1$.
As these windows move to later times, they average over an increasing number of observation rounds.
In this averaged sense, the numerical results suggest that the eigenvalues decay faster than the bounds and that their decay rates converge to $\beta$ from above; that is, over the simulated horizon we witness a transient in the slope trajectory, which has not yet settled to its asymptotic value.
Normalizing the incoming observations and shortening the path to the root both accelerate learning, in accordance with Table~\ref{tab:parameter values}.

The proof of Lemma~\ref{lem:order of contact} suggests why the transient persists. 
The bound~(8) keeps only the leading order of $\varphi_t^r$ in $q_t$: on every relay, the term $\smash{\varphi_t^j}$ and the factor $\smash{(1+q_t/\varphi_t^j)^2}$ are absorbed into its constants, as are the precisions of all children off a shortest directed path. 
While $q_t$ is not yet small, these higher-order terms remain significant and the averaged decay can be faster than its asymptotic rate. Their influence disappears slowly, making the transient longer the smaller $\beta$ is. 
The study of the transient properties is left for future work.

\subsection{Discussion}
The practical relevance of these rates depends on the time scale over which the state remains approximately constant: if learning is substantially slower, consistency is of limited benefit \citep{vives1993fast}.
Our results point to three remedies, which concern the topology of the network, the protocol, and the use of forgetting.

First, since $\beta$ depends on $(\mathcal{V},\mathcal{E})$ only through $\delta_r$, reducing this distance improves the exponent, whereas agents or branches that leave it unchanged affect only the constants.
In particular, letting agents in $\mathcal{V}\setminus\mathcal{L}$ observe the signal directly adds, in the language of Remark~\ref{re:electrical networks}, a parallel branch of constant resistance, which keeps $\varphi_t^i$ bounded as $q_t$ vanishes.
We expect our analysis to extend to this case if the observation noises remain orthogonal.
This mechanism parallels the perfectly informed agents in \citet{vives1993fast}, who place a constant weight on their private signals and thereby restore the standard rate.
Second, \citet{krishnamurthy2024slow} consider a scalar state and a path from the signal to the root, and recover the optimal rate $\Theta(t^{-1})$ by inflating the error covariance broadcast by the root, proportionally to time, before the other agents use it; the root itself retains the unscaled covariance, without which consistency is lost.
Since our exponent is determined by a shortest directed path from a leaf to the root, we expect a similar rescaling to be effective on trees, at the expense of the local optimality in Proposition~\ref{prop:optimality}.
Third, inflating $\mathbf{P}_{t-1}$ through a forgetting factor discounts past information and maintains sensitivity to new observations.
If $q_t$ thereby remains bounded above and away from zero, then so does $\varphi_t^r$, and every round contributes a nonvanishing amount of information.
Asymptotic consistency is then traded for the ability to track a time-varying state.

\section{Conclusions}\label{sec:conclusions}
We have studied recursive estimation in hierarchical networks where only the leaves observe the signal and every agent defers to the root.
Although each local update is optimal for the information it retains, the network learns slowly: the error covariance decays as $t^{-\beta}$, where $\beta$ is at least halved by every relay on a shortest path from a leaf to the root.
Hierarchies thus pay for authority with their rate of learning, irrespective of their size.
Future work will address the transient behavior, remedies to slow learning, and general graphs, where relayed noises are correlated.

\bibliography{ifacconf}

\section*{APPENDIX}\label{app:appendix}

{\textbf{Proof of Propositions~\ref{prop:observation model at edges} and~\ref{prop:optimality}.}
    Both claims are proved at once, by nested inductions: first on $t\in\mathbb{N}\cup\{0\}$ and then on the partial order defined by $(\mathcal{V},\mathcal{E})$, for a fixed $t\in\mathbb{N}$.

    We define $\mathcal{G}_{t}\coloneqq\operatorname{span}(\{\mathbf{1},\vect{x}\}\cup\{\smash[t]{e_{\tau}^{vu}}\colon (u,v)\in\smash[t]{\bar{\mathcal{E}}},\,1\leq\tau\leq t\})$ for readability, so $\smash[t]{e_{t}^{ij}\perp\mathcal{G}_{t-1}}$ for every $\smash[t]{(j,i)\in\bar{\mathcal{E}}}$ and $t\in\mathbb{N}$, by Assumption~\ref{assum:orth}.
    Alongside both propositions, we prove that
    $\smash[t]{\est{x}_{t}=\Ehat[\vect{x}\cond\mathcal{F}_{t}]}$ with $\mathcal{F}_{t} \subseteq \mathcal{G}_{t}$ for every $t\in\mathbb{N}\cup\{0\}$, and that $\smash[t]{w_{t}^{ij}}$ is a deterministic linear combination of $\smash[t]{e_{t}^{ij}}$ and of all noise terms whose edges belong to a path from $0$ to $j$.

    \emph{Base case.} At $t=0$, $\mathcal{F}_{0}=\operatorname{span}(\{\mathbf{1}\}) \subseteq\mathcal{G}_{0}$, so $\est{x}_{0} = \mathbb{E}[\vect{x}] = \smash{\Ehat[\vect{x}\cond\mathcal{F}_{0}]}$ and $\matr{P}_{0} = \mathbb{E}\bigl[(\vect{x} - \est{x}_{0})(\vect{x} - \est{x}_{0})^{\top}\bigr] \succ \matr{0}$ deterministic.

    \emph{Induction step.}
    Fix $t\in\mathbb{N}$, and suppose that Propositions~\ref{prop:observation model at edges} and~\ref{prop:optimality} hold up to and including time $t-1$.
    In particular, $\vect{x}-\est{x}_{t-1}\perp\mathcal{F}_{t-1}$ and $\smash[t]{\matr{P}_{t-1}}\succ\matr{0}$, so $q_{t}>0$.
    Furthermore, $\mathcal{G}_{t-1}\supseteq\mathcal{F}_{t-1} +\operatorname{span}(\{\vect{x}\}\cup\{\smash[t]{w_{\tau}^{vu}}\colon (u,v)\in\smash[t]{\bar{\mathcal{E}}},\,1\leq\tau < t\})$.
    By Assumption~\ref{assum:orth} then it follows that $\smash[t]{w_{t}^{ij}\perp\mathcal{G}_{t-1}}$: in particular we have $\smash[t]{\mathbb{E}[w_{t}^{ij}]=0}$, $\smash[t]{w_{t}^{ij}\perp\mathcal{F}_{t-1}}$, $\smash[t]{w_{t}^{ij}\perp\vect{x}}$, $\smash[t]{w_{t}^{ij}\perp\vect{x}-\est{x}_{t-1}}$, and $\smash[t]{w_{t}^{ij} \perp w_{\tau}^{ij'}}$ for every $\smash[t]{j'\in\bar{\mathcal{C}}_{i}}$ and $\tau < t$.

    \emph{Base case.}
    For $i\in\mathcal{L}$, the observation $\smash[t]{y_{t}^{i0} = s_{t}(\vect{x}) + e_{t}^{i0}}$ is of the form~\eqref{eq:obs at edge ansatz}, with the parameters prescribed by~(i).
    Property~(ii) is also met, since $\smash[t]{a_{t}^{i0} = 1}$ and $\smash[t]{\varphi_{t}^{i0} = \varepsilon^{i0} > 0}$ are known constants and $\smash[t]{b_{t}^{i0} = 0 \in \mathcal{F}_{t-1}}$.
    Finally, $\smash[t]{w_{t}^{i0}=e_{t}^{i0}}$ is a linear function of $\smash[t]{e_{t}^{i0}}$ alone, thus implying~(iii).

    \emph{Induction step.}
    Fix $i\in\mathcal{V}\setminus\mathcal{L}$ and suppose that~(i)--(iii) hold for every $\smash[t]{j\in\bar{\mathcal{C}}_{i}}$, that is, for every $\smash[t]{(\ell,j)\in\bar{\mathcal{E}}}$ such that $\smash[t]{\ell\in\bar{\mathcal{C}}_{j}}$.
    Suppose also that each $\smash[t]{w_{t}^{j\ell}}$ is a deterministic linear combination of $\smash[t]{e_{t}^{j\ell}}$ and all noise terms whose edges belong to a directed path from $0$ to $\ell$.
    Fix $\smash[t]{j\in\bar{\mathcal{C}}_{i}}$.
    Substituting~\eqref{eq:obs at node ansatz} into~\eqref{eq:RLS update state estimate} at $j$, premultiplying by $\smash[t]{a_{t}^{j}\vect{c}_{t}^{\top}}$, and finally substituting into~\eqref{eq:obs at edge true} yields
        \begin{equation*}
            y_{t}^{ij}=\frac{a_{t}^{j}q_{t}}{q_{t}+\varphi_{t}^{j}}
            \bigl(s_{t}(\vect{x})+w_{t}^{j}\bigr)
            +\frac{a_{t}^{j}\varphi_{t}^{j}}{q_{t}+\varphi_{t}^{j}}\,
            s_{t}(\est{x}_{t-1})+e_{t}^{ij},
        \end{equation*}
    which is~\eqref{eq:obs at edge ansatz} with the parameters in~(i).
    By~(ii) on the edges into $j$, $\smash[t]{\varphi_{t}^{j}}$ and $\smash[t]{a_{t}^{j}=(\varphi_{t}^{j})^{\eta-1}}$ are positive, deterministic, and computable offline.
    Together with $q_{t}$, so are $\smash[t]{a_{t}^{ij}}$ and $\smash[t]{\varphi_{t}^{ij}}$, while $\smash[t]{b_{t}^{ij}}$ is a multiple of $s_{t}(\est{x}_{t-1})\in\mathcal{F}_{t-1}$, thus proving~(ii).
    As $(\mathcal{V},\mathcal{E})$ is a tree, distinct edges into $i$ share no upstream edge.
    Therefore $\smash[t]{w_{t}^{ij}\perp w_{t}^{ij'}}$ for $\smash[t]{j'\neq j}$ and $\smash[t]{e_{t}^{ij}\perp w_{t}^{j}}$ so
        \begin{align*}
            \mathbb{E}[(w_{t}^{j})^{2}]
            &=(\varphi_{t}^{j})^{2}\sum_{\ell\in\bar{\mathcal{C}}_{j}}
            \frac{\mathbb{E}[(w_{t}^{j\ell})^{2}]}{(\varphi_{t}^{j\ell})^{2}}
            =\varphi_{t}^{j},\\
            \mathbb{E}[(w_{t}^{ij})^{2}]
            &=\mathbb{E}[(w_{t}^{j})^{2}]
            +\frac{\mathbb{E}[(e_{t}^{ij})^{2}]}{(a_{t}^{ij})^{2}}
            =\varphi_{t}^{j}+\frac{\varepsilon^{ij}}{(a_{t}^{ij})^{2}}
            =\varphi_{t}^{ij}.
        \end{align*}
    This proves~(iii).

    We now prove Proposition~\ref{prop:optimality} for an arbitrary $i\in\mathcal{V}$.
    By~\eqref{eq:obs at edge ansatz}, $(\smash[t]{y_{t}^{ij}}-\smash[t]{b_{t}^{ij}})/\smash[t]{a_{t}^{ij}}-s_{t}(\est{x}_{t-1}) =\smash[t]{\vect{c}_{t}^{\top}}(\vect{x}-\est{x}_{t-1})+\smash[t]{w_{t}^{ij}}
    \perp\mathcal{F}_{t-1}$. Since $\smash[t]{a_{t}^{ij}}\neq0$ is deterministic and
    $\smash[t]{b_{t}^{ij}},s_{t}(\est{x}_{t-1})\in\mathcal{F}_{t-1}$, it follows that $\smash[t]{\mathcal{F}_{t}^{i}}=\mathcal{F}_{t-1}\oplus\operatorname{span}(\{\smash[t]{\vect{c}_{t}^{\top}}(\vect{x}-\est{x}_{t-1})+\smash[t]{w_{t}^{ij}}\colon
    j\in\smash[t]{\bar{\mathcal{C}}_{i}}\})$.
    By~\eqref{eq:obs at node ansatz} and~\eqref{eq:RLS update state estimate},
    $\smash[t]{\est{x}_{t}^{i}}=\est{x}_{t-1}+\smash[t]{\vect{k}_{t}^{i}}
    (\smash[t]{\vect{c}_{t}^{\top}}(\vect{x}-\est{x}_{t-1})+\smash[t]{w_{t}^{i}})$ with
    $\smash[t]{\vect{k}_{t}^{i}}\coloneqq\matr{P}_{t-1}\vect{c}_{t}/(q_{t}+\smash[t]{\varphi_{t}^{i}})$, where $\smash[t]{\vect{c}_{t}^{\top}}(\vect{x}-\est{x}_{t-1})+\smash[t]{w_{t}^{i}}$ is the inverse-variance average of the terms $\smash[t]{\vect{c}_{t}^{\top}}(\vect{x}-\est{x}_{t-1})+\smash[t]{w_{t}^{ij}}$.
    Hence $\smash[t]{\est{x}_{t}^{i}}\in\smash[t]{\mathcal{F}_{t}^{i}}$ and $\vect{x}-\smash[t]{\est{x}_{t}^{i}}\perp\mathcal{F}_{t-1}$.
    Also, $\smash[t]{w_{t}^{ij}}\perp\vect{x}-\est{x}_{t-1}$ and, by~\eqref{eq:obs at node ansatz} and~(iii), $\mathbb{E}[\smash[t]{w_{t}^{i}}\smash[t]{w_{t}^{ij}}]=\smash[t]{\varphi_{t}^{i}}$.
    So for every $j\in\smash[t]{\bar{\mathcal{C}}_{i}}$, $\mathbb{E}\bigl[(\vect{x}-\smash[t]{\est{x}_{t}^{i}})
    (\smash[t]{\vect{c}_{t}^{\top}}(\vect{x}-\est{x}_{t-1})+\smash[t]{w_{t}^{ij}})\bigr]
    =\matr{P}_{t-1}\vect{c}_{t}-\smash[t]{\vect{k}_{t}^{i}}(q_{t}+\smash[t]{\varphi_{t}^{i}})
    =\matr{0}$.
    Therefore $\vect{x}-\smash[t]{\est{x}_{t}^{i}}\perp\smash[t]{\mathcal{F}_{t}^{i}}$, that is, $\smash[t]{\est{x}_{t}^{i}}=\smash[t]{\Ehat}[\vect{x}\cond\smash[t]{\mathcal{F}_{t}^{i}}]$.
    In particular, we have $\vect{x}-\smash[t]{\est{x}_{t}^{i}}\perp
    \smash[t]{\vect{c}_{t}^{\top}}(\vect{x}-\est{x}_{t-1})+\smash[t]{w_{t}^{i}}$, so
        \begin{align*}
            \mathbb{E}\bigl[(\vect{x}-\est{x}_{t}^{i})
            (\vect{x}-\est{x}_{t}^{i})^{\top}\bigr]
            &=\mathbb{E}\bigl[(\vect{x}-\est{x}_{t}^{i})
            (\vect{x}-\est{x}_{t-1})^{\top}\bigr]\\
            &=\matr{P}_{t-1}-\vect{k}_{t}^{i}\vect{c}_{t}^{\top}\matr{P}_{t-1}
            =\matr{P}_{t}^{i}.
        \end{align*}
    By the matrix-inversion lemma \citep[Sec.~0.7.4]{horn2012matrix} we obtain $\smash{\matr{P}^i_t = (\matr{P}_{t-1}^{-1} + \vect{c}_t\vect{c}_t^\top/\varphi^i_t)^{-1}}$, which is positive definite, deterministic, and computable offline.

    At the root, $\smash[t]{\mathcal{F}_{t}\subseteq\mathcal{F}_{t}^{r}}$, since $\smash[t]{\mathcal{F}_{t-1}\subseteq\mathcal{F}_{t}^{r}}$ and $\smash[t]{\est{x}_{t}=\est{x}_{t}^{r}\in\mathcal{F}_{t}^{r}}$.
    Hence $\smash[t]{\est{x}_{t}=\Ehat[\vect{x}\cond\mathcal{F}_{t}]}$ and $\smash[t]{\matr{P}_{t}=\matr{P}_{t}^{r}\succ\matr{0}}$.
    Finally, $\mathcal{F}_{t}\subseteq\smash[t]{\mathcal{F}_{t}^{r}}\subseteq\mathcal{G}_{t}$, completing the induction on $t$.
\hfill$\square$}

{\textbf{Proof of Lemma~\ref{lem:order of contact}.}
    Let $\gamma\coloneqq4 - 2\eta$ and $\alpha\colon\mathbb{N}\cup\{0\}\to\mathbb{R}$ be given by $\alpha(\xi)\coloneqq2(\gamma^{\xi}-1)/(\gamma-1)$.
    We show by induction on the partial order defined by $(\mathcal{V},\mathcal{E})$ that $\varphi_{t}^{i}=\Theta\bigl(\smash[t]{q_{t}^{-\alpha(\delta_{i})}}\bigr)$.

    \emph{Base case.}
    If $i\in\mathcal{L}$, then $\delta_{i} = 0$, $\alpha(\delta_{i}) = 0$, and $\varphi_{t}^{i} = \Theta(1)$.

    \emph{Induction step.}
    Fix $i\in\mathcal{V}\setminus\mathcal{L}$ and assume that the statement holds for every $j\in\bar{\mathcal{C}}_{i}$.
    By Assumption~\ref{assum:pe} and~\eqref{eq:RLS update variance}, $q_{t}\leq\bar{c}\operatorname{tr}(\matr{P}_{0})$ for every $t\in\mathbb{N}$.
    Since $q_{t}$ is bounded, the induction hypothesis implies that $\smash[t]{\varphi_{t}^{j}}$ is bounded away from zero, and hence $\smash[t]{1+q_{t}/\varphi_{t}^{j} = \Theta(1)}$.
    Proposition~\ref{prop:observation model at edges}(i) and~\eqref{eq:obs at node ansatz} yield
        \begin{equation*}
            \begin{aligned}
                \varphi_{t}^{ij}
                & = \varphi_{t}^{j} + \frac{\varepsilon^{ij}(q_{t} + \varphi_{t}^{j})^{2}}{(a_{t}^{j}q_{t})^{2}}
                = \varphi_{t}^{j} + \frac{\varepsilon^{ij}(\varphi_{t}^{j})^{\gamma}}{q_{t}^{2}} \biggl(1 + \frac{q_{t}}{\varphi_{t}^{j}}\biggr)^{2} \\
                & = \Theta\bigl(q_{t}^{-\alpha(\delta_{j})}\bigr) + \Theta\bigl(q_{t}^{-\alpha(\delta_{j}+1)}\bigr)
                = \Theta\bigl(q_{t}^{-\alpha(\delta_{j}+1)}\bigr),
            \end{aligned}
        \end{equation*}
    where membership follows from the induction hypothesis and the identity
    $\gamma\,\alpha(\xi)+2=\alpha(\xi+1)$, and the equality from $\alpha(\delta_{j}+1)>\alpha(\delta_{j})$, as $q_{t}$ is bounded.
    Finally,
        \begin{equation*}
            \frac{1}{\varphi_{t}^{i}}=\sum_{j\in\bar{\mathcal{C}}_{i}}\frac{1}{\varphi_{t}^{ij}} =
            \Theta\bigl(q_{t}^{\min_{j}\alpha(\delta_{j}+1)}\bigr)=\Theta\bigl(q_{t}^{\alpha(\delta_{i})}\bigr),
        \end{equation*}
    as $\alpha$ is strictly increasing and $\delta_{i}=1+\min_{j}\delta_{j}$.
\hfill$\square$}

{\textbf{Proof of Theorem~\ref{thm:rate of convergence}.}
    Since $\matr{P}_{t-1} \succ \matr{0}$, the eigenvalues of $\matr{P}_{t-1}$ are the reciprocals of the eigenvalues of $\smash{\matr{P}_{t-1}^{-1}}$.
    Hence, \eqref{eq:bound on error covariance} holds if and only if
        \begin{align*}
            \lambda_{\min}(\matr{P}_{t-1}^{-1}) & = \Omega\bigl((t+1)^{\beta}\bigr), \\
            \lambda_{\max}(\matr{P}_{t-1}^{-1}) & = \mathcal{O}\bigl((t+1)^{\beta}\bigr),
        \end{align*}
    which correspond to the upper and lower bounds in \eqref{eq:bound on error covariance}, respectively.
    Two facts are used for proving both bounds.
    First, inverting \eqref{eq:RLS info form single agent} and bounding $\smash[t]{\varphi_{t}^{r}}$ by Lemma~\ref{lem:order of contact} yields
        \begin{equation}\label{eq:Pinv-bounds}
            \matr{P}_{t-1}^{-1} + \frac{q_{t}^{\alpha_{r}}}{\bar{\varphi}}\, \vect{c}_{t} \vect{c}_{t}^{\top}
            \preceq \matr{P}_{t}^{-1} \preceq
            \matr{P}_{t-1}^{-1} + \frac{q_{t}^{\alpha_{r}}}{\underaccent{\bar}{\varphi}}\, \vect{c}_{t} \vect{c}_{t}^{\top}
        \end{equation}
    for every $t \in \mathbb{N}$.
    Second, the left inequality in \eqref{eq:Pinv-bounds} implies $\{\smash[t]{\matr{P}_{t}^{-1}}\colon t \in \mathbb{N} \cup \{0\}\}$ is nondecreasing in the L\"owner order.

\emph{Upper bound.}
    We first bound the growth of $\smash[t]{\boldsymbol{\xi}^{\top} \matr{P}_{t}^{-1} \boldsymbol{\xi}}$ over a time window of $\bar{t} + 1$ time steps.
    Fix $t \in \mathbb{N}$ and a unit vector $\boldsymbol{\xi} \in \mathbb{R}^{n}$, and let $\tau \in \{t, \dots, t + \bar{t}\}$.
    The Cauchy--Schwarz inequality and monotonicity give
        \begin{align}
            (\boldsymbol{\xi}^{\top} \vect{c}_{\tau})^{2}
            &= (\boldsymbol{\xi}^{\top} \matr{P}_{\tau-1}^{-1/2} \matr{P}_{\tau-1}^{1/2} \vect{c}_{\tau})^{2}
            \le \|\matr{P}_{\tau-1}^{-1/2} \boldsymbol{\xi}\|^{2}_{2} \|\matr{P}_{\tau-1}^{1/2} \vect{c}_{\tau}\|^{2}_{2} \nonumber\\
            &= \bigl(\boldsymbol{\xi}^{\top} \matr{P}_{\tau-1}^{-1} \boldsymbol{\xi}\bigr)\, q_{\tau}
            \le \bigl(\boldsymbol{\xi}^{\top} \matr{P}_{t+\bar{t}}^{-1} \boldsymbol{\xi}\bigr)\, q_{\tau} .
            \label{eq:CS}
        \end{align}
    Combining the left inequality in \eqref{eq:Pinv-bounds}, evaluated at time $\tau$, with the bound $q_{\tau} \ge \smash[t]{(\boldsymbol{\xi}^{\top} \vect{c}_{\tau})^{2}} / (\smash[t]{\boldsymbol{\xi}^{\top}} \smash{\matr{P}_{t+\bar{t}}^{-1}} \boldsymbol{\xi})$ from \eqref{eq:CS}, gives
    \begin{equation*}
        \boldsymbol{\xi}^{\top} \bigl(\matr{P}_{\tau}^{-1} - \matr{P}_{\tau-1}^{-1}\bigr) \boldsymbol{\xi}
        \ge
        \frac{(\boldsymbol{\xi}^{\top} \vect{c}_{\tau})^{2(1+\alpha_{r})}} {\bar{\varphi}\, (\boldsymbol{\xi}^{\top} \matr{P}_{t+\bar{t}}^{-1} \boldsymbol{\xi})^{\alpha_{r}}} .
    \end{equation*}
    Consequently, summing over $\tau \in\{ t, \dots, t + \bar{t}\}$ and telescoping yields the following increment over the prescribed time window:
        \begin{equation}\label{eq:window}
            \boldsymbol{\xi}^{\top} \bigl(\matr{P}_{t+\bar{t}}^{-1} - \matr{P}_{t-1}^{-1}\bigr)
            \boldsymbol{\xi}
            \ge
            \frac{\sum_{\tau=t}^{t+\bar{t}} (\boldsymbol{\xi}^{\top} \vect{c}_{\tau})^{2(1+\alpha_{r})}} {\bar{\varphi}\, (\boldsymbol{\xi}^{\top} \matr{P}_{t+\bar{t}}^{-1} \boldsymbol{\xi})^{\alpha_{r}}}.
        \end{equation}

    We will now bound the numerator in \eqref{eq:window} from below by a strictly positive constant independent of $t$ and $\boldsymbol{\xi}$.
    Since $\alpha_{r} \ge 0$, the map $\xi \mapsto \smash[t]{\xi^{1+\alpha_{r}}}$ is convex on $[0, \infty)$. Applying Jensen's inequality to the average of $\smash[t]{(\boldsymbol{\xi}^{\top} \vect{c}_{\tau})^{2}}$ over the time window, and then Assumption~\ref{assum:pe} with $\|\boldsymbol{\xi}\|_{2} = 1$, we obtain
        \begin{align*}
            \frac{1}{\bar{t}+1} \sum_{\tau=t}^{t+\bar{t}} (\boldsymbol{\xi}^{\top} \vect{c}_{\tau})^{2(1+\alpha_{r})}
            \ge
            \bigg[ \frac{1}{\bar{t}+1} \sum_{\tau=t}^{t+\bar{t}}
            (\boldsymbol{\xi}^{\top} \vect{c}_{\tau})^{2} \bigg]^{1+\alpha_{r}} \\
            \quad = \bigg[ \frac{\boldsymbol{\xi}^{\top} \big(\sum_{\tau=t}^{t+\bar{t}}
            \vect{c}_{\tau} \vect{c}_{\tau}^{\top}\big) \boldsymbol{\xi}}{\bar{t}+1} \bigg]^{1+\alpha_{r}}
            \ge
            \bigg( \frac{\underaccent{\bar}{c}}{\bar{t}+1} \bigg)^{1+\alpha_{r}}.
        \end{align*}
    Substituting into \eqref{eq:window} yields
        \begin{equation}\label{eq:key}
            \boldsymbol{\xi}^{\top} \bigl(\matr{P}_{t+\bar{t}}^{-1} - \matr{P}_{t-1}^{-1}\bigr)
            \boldsymbol{\xi}
            \ge
            \frac{\kappa}{(\boldsymbol{\xi}^{\top} \matr{P}_{t+\bar{t}}^{-1} \boldsymbol{\xi})^{\alpha_{r}}},
        \end{equation}
    where $\kappa \coloneqq \smash[t]{\underaccent{\bar}{c}^{\,1+\alpha_{r}}} / (\bar{\varphi}\, \smash[t]{(\bar{t}+1)^{\alpha_{r}}})$.

    It remains to turn \eqref{eq:key} into a recursion for $\lambda_{\min}(\smash[t]{\matr{P}_{t}^{-1}})$.
    Let $\boldsymbol{\xi}$ be a unit eigenvector associated with $\lambda_{\min}(\smash{\matr{P}_{t+\bar{t}}^{-1}})$.
    Since $\smash{\boldsymbol{\xi}^{\top} \matr{P}_{t-1}^{-1} \boldsymbol{\xi}} \ge \lambda_{\min}(\smash{\matr{P}_{t-1}^{-1}})$, multiplying both sides of \eqref{eq:key} by $\smash[t]{\lambda_{\min}^{\alpha_{r}}}(\smash{\matr{P}_{t+\bar{t}}^{-1}})$ and using monotonicity yields
        \begin{align}
            \lambda_{\min}^{1+\alpha_{r}}(\matr{P}_{t+\bar{t}}^{-1})
            &\ge \lambda_{\min}(\matr{P}_{t-1}^{-1})\,
                 \lambda_{\min}^{\alpha_{r}}(\matr{P}_{t+\bar{t}}^{-1}) + \kappa
            \nonumber\\
            &\ge \lambda_{\min}^{1+\alpha_{r}}(\matr{P}_{t-1}^{-1}) + \kappa .
            \label{eq:recursion}
        \end{align}
    Therefore, $\smash[t]{\lambda_{\min}^{1+\alpha_{r}}(\matr{P}_{\tau}^{-1})}$ grows by at least $\kappa$ every $\bar{t} + 1$ time steps.
    Applying \eqref{eq:recursion} $\lfloor (t-1)/(\bar{t}+1) \rfloor$ times backward from $\tau = t - 1$ and then using monotonicity, we get
        \begin{equation*}
            \lambda_{\min}^{1+\alpha_{r}}(\matr{P}_{t-1}^{-1})
            \ge \lambda_{\min}^{1+\alpha_{r}}(\matr{P}_{0}^{-1})
            + \kappa \bigg\lfloor \frac{t-1}{\bar{t}+1} \bigg\rfloor
            = \Omega(t+1),
        \end{equation*}
    since $\smash[t]{\matr{P}_{0}^{-1}} \succ \matr{0}$.
    Taking the $(1+\alpha_{r})$-th root gives the upper bound in \eqref{eq:bound on error covariance}, namely $\lambda_{\min}(\smash{\matr{P}_{t-1}^{-1}}) = \smash{\Omega\bigl((t+1)^{\beta}\bigr)}$.

\emph{Lower bound.}
    We now use the upper bound to control the growth of $\operatorname{tr}(\smash[t]{\matr{P}_{t}^{-1}})$, which dominates $\lambda_{\max}(\smash[t]{\matr{P}_{t}^{-1}})$ since $\smash[t]{\matr{P}_{t}^{-1}} \succ \matr{0}$.
    Taking the trace of the right inequality in~\eqref{eq:Pinv-bounds} at time $\tau$, using $\operatorname{tr}(\vect{c}_{\tau}\smash[t]{\vect{c}_{\tau}^{\top}}) = \smash[t]{\|\vect{c}_{\tau}\|_{2}^{2}} \le \bar{c}$ by Assumption~\ref{assum:pe}, summing over $\tau \in \{1, \dots, t-1\}$, and telescoping, we get
        \begin{equation}\label{eq:trace-sum}
            \operatorname{tr}(\matr{P}_{t-1}^{-1})
            \le \operatorname{tr}(\matr{P}_{0}^{-1}) + \frac{\bar{c}}{\underaccent{\bar}{\varphi}} \sum_{\tau=1}^{t-1} q_{\tau}^{\alpha_{r}}.
        \end{equation}
        Since $q_{\tau} \le \smash[t]{\|\vect{c}_{\tau}\|_{2}^{2}}\, \lambda_{\max}(\matr{P}_{\tau-1})$ and $\smash[t]{\|\vect{c}_{\tau}\|_{2}^{2}} \le \bar{c}$, then
        \begin{equation*}
            q_{\tau}^{\alpha_{r}}
            \le \frac{\bar{c}^{\,\alpha_{r}}}{\lambda_{\min}^{\alpha_{r}}(\matr{P}_{\tau-1}^{-1})}
            = \mathcal{O}\bigl((\tau+1)^{-\alpha_{r}\beta}\bigr).
        \end{equation*}
    Moreover, as $\xi \mapsto \smash[t]{\xi^{-\alpha_{r}\beta}}$ is nonincreasing on $(0, \infty)$, each $\smash[t]{(\tau+1)^{-\alpha_{r}\beta}}$ is at most the integral of $\smash[t]{\xi^{-\alpha_{r}\beta}}$ over $[\tau, \tau+1]$.
    Since also ${1 - \alpha_{r}\beta = \beta}$, it follows that
        \begin{equation*}
            \sum_{\tau=1}^{t-1} (\tau+1)^{-\alpha_{r}\beta}
            \le \int_{1}^{t} \xi^{-\alpha_{r}\beta}\, \mathrm{d}\xi
            = \frac{t^{\beta} - 1}{\beta}
            = \mathcal{O}\bigl((t+1)^{\beta}\bigr).
        \end{equation*}
    Combining with~\eqref{eq:trace-sum}, we obtain $\lambda_{\max}(\smash{\matr{P}_{t-1}^{-1}}) \le \operatorname{tr}(\smash{\matr{P}_{t-1}^{-1}}) = \smash{\mathcal{O}\bigl((t+1)^{\beta}\bigr)}$, thus proving the lower bound in~\eqref{eq:bound on error covariance}.
\hfill$\square$}

\smallskip {}

\end{document}